\documentclass[prd,twocolumn,floatfix,amsmath,nofootinbib,amssymb,floatfix]{revtex4}
\usepackage{graphicx,color,dcolumn,booktabs,bm}
\usepackage{longtable,lscape}
\usepackage{txfonts}
\usepackage{overpic}
\usepackage{amssymb}
\usepackage{indentfirst}
\usepackage{feynmf}   %{feynmp}
\usepackage{slashed}  %for Feynman symbols
\usepackage{cases}
\usepackage{color}
\usepackage{multirow}
\usepackage{enumerate}
\usepackage{graphicx,color,dcolumn,booktabs,bm}
\usepackage{epsfig}
\usepackage[colorlinks, citecolor=blue,anchorcolor=red,menucolor=red, linkcolor=red,filecolor=red,urlcolor=blue,frenchlinks=red]{hyperref}
\newcommand{\vew}{\upsilon_{EW}}
\newcommand{\vso}{\upsilon_{1}}
\newcommand{\vst}{\upsilon_{2}}
\graphicspath{{Figures/}} %

\begin{document}

\title{Naturalness Implications within the Two-Real-Scalar-Singlet beyond the SM}
\author{Jamal Ou aali$^{1}$}\email{j.ouaali96@gmail.com}
\author{Bouzid Manaut$^{1}$}\email{bmanaut@gmail.com}
\author{Larbi Rahili$^{2}$\footnote{Corresponding author}}\email{rahililarbi@gmail.com}
\author{Souad Semlali$^{3}$}\email{s.seemlali@gmail.com}
\affiliation{
$^1$ERPTM, Facult\'e Polydisciplinaire, Universit\'e Sultan Moulay Slimane, B\'eni Mellal, Morocco\\
$^2$EPTHE, Faculty of Sciences, Ibn Zohr University, B.P 8106, Agadir, Morocco\\
$^3$LPFAS, Facult\'e Polydisciplinaire de Safi, Sidi Bouzid, BP 4162, Safi, Morocco.
}

\begin{abstract}
The aim of this study is to investigate the quadratic divergences using dimensional regularization within the context of the standard model extended by two real scalar singlets (TRSM). This extension provides three neutral scalar fields that mix, after developing its {\it vev}'s, leading to three CP-even Higgs bosons, namely: $h_1$, $h_2$ and $h_3$, which would offer a wide phenomenology at the Large Hadron Collider (LHC) as reported recently. Furthermore, to fulfill the Veltman conditions for those three fields, we dwell on the one-loop level ($d_L=2$) of dimensional regularization calculations, assuming $R_{\xi}$ Feynman-'t Hooft gauge-invariant generalization. We show that the divergence cancelation could take place in the framework of the TRSM, 
and thereby predicts a stringent constraint on the space parameters as well as the new physics (NP) scale, and yet remain consistent with current experimental measurements at 13 TeV. 

\end{abstract}
\date{\today}
\maketitle

\section{ Introduction }
\label{sec:intro}
From the outset, the quadratic divergencies cancellation was and remains important in assisting our understanding of fundamental particle physics. The first accurate study of the subject in the standard model (SM) was initially hypothesized by Decker and Pestiean \cite{Decker:1979nk} before being elaborated by Veltman \cite{Veltman:1980mj}, in which they demonstrated that the contributions to the quadratic divergence of the Higgs mass arising from both fermionic and bosonic sectors could cancel, providing thus bounds on the free parameters of the model. And, now that the Higgs boson is well established by LHC experiments \cite{Aad:2012tfa,Chatrchyan:2012ufa} with more precise measurements \cite{Sopczak:2020vrs}, as well as no other new particles, have yet materialized at the LHC,
any addressed extension beyond SM (BSM) must take this criterion into account to continue the search for new physics. Indeed, utilizing this cancellation as an additional constraint reveals the value of including a specific multiplet into the SM sector that matches the existing restrictions.

In the literature, two approaches are usually adopted for deriving the so-called Veltman condition (VC), either by using dimensional regularization \cite{Einhorn:1992um}, which is the most appreciated regarding the gauge and Lorentz invariances that respects, or through a cut-off regularization scheme \cite{Oleszczuk:1994st}. According to the first concept, the SM Higgs boson receives, at one-loop, significant quantum corrections on its mass which reads as,
\begin{equation}
\delta\,m_{H}^2 \propto m_H^2+m_Z^2+2m_W^2-4m_t^2,
\label{dm_SM}
\end{equation}
yielding a Higgs mass of $\approx\,314$ GeV which is in stark contradiction with experimental data, so then new physics is required to scale up the SM contribution. 

In fact, debates over naturalness were an open question for physicists for a long time \cite{Pivovarov:2007dj}, and many extensions BSM have already touched upon this purpose either in the SM extended with real singlet \cite{Grzadkowski:2009bp, Karahan:2014ola}, two Higgs doublet model \cite{Darvishi:2017bhf} and beyond \cite{Ait-Ouazghour:2020slc}, or SM+triplet \cite{Rahili:2015nel, Rahili:2018ert}, thus attempting to resolve the fine-tuning problem by applying the VC criterion. And to us, we underline an extension where two real scalar singlets are added to SM sector, TRSM. In fact, this latter has been extensively examined BSM, notably to provide a basis for reopening the dark matter arguing \cite{Abada:2011qb,Ahriche:2012ei} and a wide phenomenological context \cite{Ahriche:2013vqa,Robens:2019kga}. However, it remains to be seen whether the VC constraints implementation could be used are compatible with theoretical requirements and experimental data.

This work is outlined as follows: in section \ref{sec:model}, we briefly review the main features of TRSM and present the full set of constraints on the parameters of the Higgs potential. Section \ref{sec:vc-pheno} is devoted to the derivation of the Veltman conditions in TRSM, the analysis, and discussion of the results performed, with emphasis on the effects of VC on the model parameters. Conclusion with a summary of our results will be drawn in section \ref{sec:conclusion}.

\section{The TRSM Model}
\label{sec:model}
\subsection{Scalar Potential \& Higgs Masses}
In order to look for a wide variety of phenomenological processes, the TRSM postulates extending the SM Higgs sector by two real scalar field $S_1$ and $S_2$ which transform under the discrete symmetry $\mathbb{Z}_{2}^{0}\otimes\mathbb{Z}_{2}^{1}$ as 
\begin{equation}
\begin{split}
\mathbb{Z}_{2}^{(0)} :\quad (S_{1},S_{2}) \rightarrow (-S_{1},S_{2}) \\
\mathbb{Z}_{2}^{(1)} :\quad (S_{1},S_{2}) \rightarrow (S_{1},-S_{2})
\end{split}
\end{equation} 
Both singlets do not contribute to gauge interactions, which would explain its partial derivative in the most general, renormalizable, and symmetric Lagrangian density, that reads as,
\begin{equation}
\mathcal{L} = (D_{\mu}H)^{\dagger}(D^{\mu}H) + \frac{1}{2} \sum_{i=1}^{2} \partial_{\mu}S_{i}\partial^{\mu}S_{i} - V(H,S_1,S_2)
\end{equation}
where 
\begin{align}
V(H,S_i) &= \mu_{H}^{2}H^{\dagger}H + \lambda_H \big( H^{\dagger}H \big)^{2} + \sum_{i=1}^{2} \big(\mu_{S_i}^{2}S_{i}^{2} + \frac{\lambda_i}{2}S_{i}^{4} \big) \nonumber\\
& + \lambda_{3} \big(S_{1} S_{2} \big)^{2} + H^{\dagger}H\,\big(\lambda_{4}S_{1}^{2} + \lambda_{5}S_{2}^{2} \big)
\label{eq1:pot}
\end{align}
is the scalar potential, and $H^T=\big(\Phi^+, \Phi^0\big)$ stands for the SM Higgs doublet. It is worth mentioning here that the charged ($\Phi^+$) and imaginary-part ($\Im[\Phi^0]$) are the Goldstone bosons eaten by the longitudinal components of W and Z boson, respectively. 

After breaking the electroweak symmetry (EWSB), the $\Phi^0$, $S_1$ and $S_2$ neutral components get non-vanishing {\it vev}, respectively given by $\vew$ (=246 GeV), $\vso$ and $\vst$. 
\begin{equation}
\langle H \rangle = \begin{pmatrix}
0 \\
\frac{\vew}{\sqrt{2}}
\end{pmatrix}, \quad \langle S_1 \rangle = \frac{\vso}{\sqrt{2}} \quad {\rm and} \quad \langle S_2 \rangle = \frac{\vst}{\sqrt{2}}
\end{equation}
and then, three minimization equations read as
\begin{equation}
\begin{aligned}
-2\,\mu_{H}^{2}  &= 2 \lambda_H \vew^{2} + \lambda_4 \vso^{2} + \lambda_5 \vst^{2} \\
-2\,\mu_{S_1}^{2}  &= \lambda_4 \vew^{2} + \lambda_1 \vso^{2} + \lambda_3 \vst^{2}  \\
-2\,\mu_{S_2}^{2}  &= \lambda_5 \vew^{2} + \lambda_3 \vso^{2} + \lambda_2 \vst^{2} 
\end{aligned}
\end{equation}
while the three non-physical fields involving the same electric charges ($Q=0$) mix each other, leading to a three massive physical CP even Higgs bosons as,
\begin{equation}
\begin{pmatrix}
h_1 \\
h_2 \\
h_3
\end{pmatrix} = R\, 
\begin{pmatrix}
\phi_H \\
\phi_1 \\
\phi_2
\end{pmatrix}.
\end{equation}
Here, $\phi_H$, $\phi_1$ and $\phi_2$ stand respectively for the real part of the neutral components $\Phi^0$, $S_1$ and $S_2$, and the $R_{\alpha_i}$ is a $3$ by $3$ unitary matrix given by,
\begin{equation}
R =
\begin{pmatrix}
c_{1} c_{2} & s_{1} c_{2} & s_{2}\\
-(c_{1} s_{2} s_{3} + s_{1} c_{3}) & c_{1} c_{3} - s_{1} s_{2} s_{3} & c_{2} s_{3} \\
- c_{1} s_{2} c_{3} + s_{1} s_{3} & -(c_{1} s_{3} + s_{1} s_{2} c_{3}) & c_{2}  c_{3}
\end{pmatrix}
\label{eq:matrix-trans}
\end{equation}
The short notations $c_i$ and $s_i$ stand respectively for $\cos\alpha_i$ and $\sin\alpha_i$, where the mixing $\alpha_i$ (i=1,2,3), all lie in the  interval $\big[-\pi/2, \pi/2\big]$. 

\noindent
At the breaking scale, the dimensionless quartic couplings ($\lambda_H$, $\lambda_1$, $\lambda_2$, $\lambda_3$, $\lambda_4$ and $\lambda_5$) can be expressed completely in terms of the doublet-singlets {\it vev}'s, as well as of the Higgs masses taken on the following sequencing, $m_{h_1} < m_{h_2} < m_{h_3}$. Indeed, by drawing on the diagonalization property $$\text{diag}(m^2_{h_1},m^2_{h_2},m^2_{h_3}) = R M^2_{S} {R^{\text T}},$$ where $M^2_{S}$ is a $3\times3$ mass matrix for the CP even Higgs sector, one gets
\begin{align}
\lambda_H &= \frac{1}{2\vew^2} \sum_{i=1}^{3} R_{i1}^2 m_{h_i}^2, &
\lambda_1 & = \frac{1}{\vso^2} \sum_{i=1}^{3} R_{i2}^2 m_{h_i}^2, \nonumber\\
\lambda_2 & = \frac{1}{\vst^2} \sum_{i=1}^{3} R_{i3}^2 m_{h_i}^2, &  
\lambda_3 & = \frac{1}{\vso\vst} \sum_{i=1}^{3} R_{i2} R_{i3} m_{h_i}^2, \nonumber\\ 
\lambda_4 & = \frac{1}{\vso\vew} \sum_{i=1}^{3} R_{i1} R_{i2} m_{h_i}^2, &
\lambda_5 & = \frac{1}{\vst\vew} \sum_{i=1}^{3} R_{i1} R_{i3} m_{h_i}^2.
\label{eq:lam_coup}
\end{align}
In our study and to assess the TRSM space parameter, a randomly scan over the Higgs masses as well as the singlets {\it vev}'s up to 1.2 $TeV$ has been performed, while requiring the light Higgs boson $h_{1}$ to behave as the SM Higgs boson observed at 125 GeV. It is in this vein that the above dimensionless couplings can be determining factors since they are subject to theoretical and experimental scrutinies as we shall see in the following.

\subsection{Theoretical \& Experimental Constraints}
\label{sec:constraints}
Theoretical requirements are often justified as being necessary to uncover the regularities hidden in the experiment data, by identifying new directions for high energy physics as simple and precise away as possible. In this context, the TRSM Higgs potential parameters are not free but have to obey several constraints arising from such requirements, and their range is set based on the following constraints:

\noindent
\underline{\it Vacuum stability}\\
It is a necessary condition for the potential to be bounded from below (BFB) when the doublet and singlets fields become large in any direction in the space. The TRSM constraints ensuring that issue read \cite{Robens:2019kga},
\begin{align}
& \lambda_H, \lambda_{1}, \lambda_{2} > 0 \nonumber\\ 
& a_{1} \equiv \xi_{1} + \sqrt{\lambda_H\lambda_{1}} > 0 \nonumber\\
& a_{2} \equiv \xi_{2} + \sqrt{\lambda_H\lambda_{2}} > 0 \nonumber\\
& a_{3} \equiv \eta + \sqrt{\lambda_{1}\lambda_{2}} > 0 \nonumber\\
& a_{4} \equiv \sqrt{\lambda_H\lambda_{1}\lambda_{2}} + \lambda_{4}\sqrt{\lambda_{2}} + \lambda_{5}\sqrt{\lambda_{1}} + \lambda_{3}\sqrt{\lambda_H} \nonumber\\
&\quad+ 2\sqrt{a_{1}a_{2}a_{3}} > 0,
\end{align}
but although highly efficient and reliable evaluation way that provides, BFB remains an insufficient condition for constraining the space parameter.

\noindent
\underline{\it Unitarity}\\
Besides, the above BFB constraints need to be supplemented by unitarity, which holds that the probability of all possible quantum interactions must add up to one. Nevertheless, at the large $\sqrt{s}$ limit, i.e. in the limit where any mass scale in the TRSM is below the energies of the scattering processes, only the 2-body of this latter arising from the quartic terms of the scalar potential have been considered. The results from \cite{Robens:2019kga} for the TRSM are summarized,
\begin{eqnarray}
&& \lvert \lambda_H \rvert < 4\pi,  \\
&& \lvert \lambda_{3} \rvert, \lvert \lambda_{4} \rvert, \lvert \lambda_{5} \rvert < 8\pi,  \\
&& \lvert x_{1,2,3} \rvert < 16\,\pi 
\end{eqnarray}
where $x_{1,2,3}$ are the three real roots of the cubic polynomial
\begin{align}
&x^3 - 3x^{2}\,\Big[4\lambda_H + (\lambda_{1} + \lambda_{2})\Big]  + x\,\Big[36 \lambda_H \big(\lambda_{1} + \lambda_{2} \big) \nonumber\\
&-4 \big(\lambda_{4}^{2} + \lambda_{5}^{2}\big) - \lambda_{3}^{2} + 9\lambda_{1}\lambda_{2}\Big] + \lambda_H\,\big(12\lambda_{3}^{2} - 108\lambda_{1}\lambda_{2}\big) \nonumber\\
&-8\lambda_{3}\lambda_{4}\lambda_{5} + 12\,\big(\lambda_{2}\lambda_{4}^{2} + \lambda_{1}\lambda_{5}^{2}\big) = 0 
\end{align}
It may be recalled that the quartic couplings involving $S_i^2$ and $S_i^4$ are of particular interest,  which is why the above constraints are closely correlated with the perturbativity; where the absolute values of all dimensionless quartic couplings must be $\le4\pi$.

\noindent
\underline{\it oblique parameters}\\
It should be pointed out, as the singlets fields do not contribute to gauge bosons masses, that there is no deviation of the $\rho$ parameter prediction within the TRSM and the SM. Thus, assuming the $U_{\rm TRSM}\approx0$, the only radiative corrections of the electroweak parameters within this model are coming from the oblique parameters $S$ and $T$, where the major contribution comes from the loops involving the new singlet-like scalars. In our study, the general definition in Ref\cite{Grimus:2007if} is adopted and in keeping with the transformation scheme given by Eq.(\ref{eq:matrix-trans}), the corresponding expressions read, 
\begin{align}
S_{\rm TRSM} &=  \frac{1}{24\pi} \bigg\{ \big(R_{11}^2+R_{21}^2-1\big) \ln(m_{h_1}^2)+\big(R_{12}^2+R_{22}^2\big) \ln(m_{h_2}^2) \nonumber\\
&+ \big(R_{13}^2+R_{23}^2\big) \ln(m_{h_3}^2) + \big( c_1^2c_2^2 -1\big) G(m_{h_1}^2,m_{Z}^2) \\
&+ s_1^2c_2^2 G(m_{h_2}^2,m_{Z}^2)+s_2^2 G(m_{h_3}^2,m_{Z}^2) \bigg\} \nonumber \\
T_{\rm TRSM} &=  \frac{3}{16\pi s_w^2m_W^2} \bigg\{ \big(1-c_1^2c_2^2\big)\Big[F(m_Z^2,m_{h_1}^2)-F(m_W^2,m_{h_1}^2)\Big] \nonumber\\
&+ s_1^2c_2^2\Big[F(m_Z^2,m_{h_2}^2)-F(m_W^2,m_{h_2}^2)\Big] \\
&+ s_2^2\Big[F(m_Z^2,m_{h_3}^2)-F(m_W^2,m_{h_3}^2)\Big] \bigg\} \nonumber
\label{eq:pot-sc}
\end{align}
where $R_{ij}$ are the matrix entries in Eq.(\ref{eq:matrix-trans}), $G\left(I,J \right)$ and $F\left(I,J \right)$ are two mathematical functions given by \cite{Grimus:2008nb}
\begin{align}
G\left( I, J \right) &=  -\frac{79}{3}+9\frac{I}{J}-2\frac{I^2}{J^2}+\Big(-10+18\frac{I}{J}-6\frac{I^2}{J^2}+\frac{I^3}{J^3} \nonumber\\
  &-9\frac{I+J}{I-J}\Big)\ln(J)+\Big(12-4\frac{I}{J}+\frac{I^2}{J^2}\Big)\frac{f(I,I^2-4IJ)}{J} \nonumber \\
F\left( I, J \right) &= 
\left\{ \begin{array}{lcl}
\displaystyle{
\frac{I + J}{2} - \frac{I J}{I - J}\, \ln{\frac{I}{J}}
}
& \Leftarrow & I \neq J,\\*[1mm]
0 & \Leftarrow & I = J.
\end{array} \right.
\end{align}
with
\begin{equation}
f \left(t,r\right) \equiv \left\{ \begin{array}{lcl}
{\displaystyle
\sqrt{r}\, \ln{\left| \frac{t - \sqrt{r}}{t + \sqrt{r}} \right|}
} & \Leftarrow & r > 0,
\\*[1mm]
0 & \Leftarrow & r = 0,
\\*[1mm]
{\displaystyle
2\, \sqrt{-r}\, \arctan{\frac{\sqrt{-r}}{t}}
} & \Leftarrow & r < 0.
\end{array} \right.
\end{equation}
\noindent
Experimentally, the global fit yields \cite{Tanabashi:2018oca}
\begin{equation}
T=0.07\pm0.06 \quad,\quad S=0.02\pm0.07
\label{eq:Tpara }
\end{equation}
and the TRSM fall safely within the $2\sigma$ allowed region of these oblique parameters.

\section{Naturalness Criteria and Phenomenology}
\label{sec:vc-pheno}
\subsection{Modified VC}
\label{subsec:mvc}
Even if the parameters introduced by the scalar potential could be determined from the requirements given above; as well as from its appearance in the couplings of the doublet and singlets fields, some strong constraints on those parameters, related to the fine-tuning problem originating from the radiative corrections to the Higgs mass, can be derived using a new argument such as the Veltman quadratic cancellation conditions. In this regard, the dimensional regularization approach would be much more consistent, and particularly suitable for preserving the local gauge symmetry of the underlying Lagrangian.
\begin{figure}[!ht]
\hspace*{-0.50cm}
\begin{minipage}{6.5cm}
\includegraphics[width=6.5cm,keepaspectratio=true]{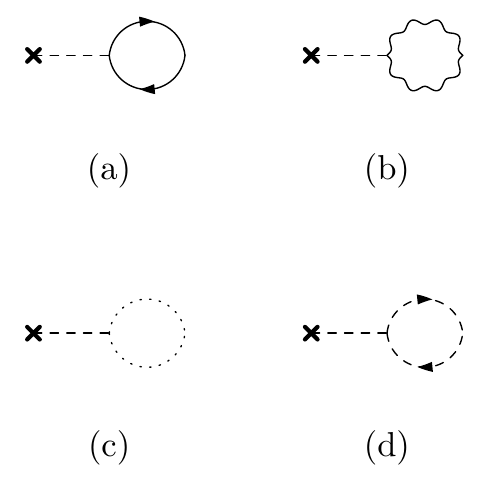}
\end{minipage}
\caption{Feynman diagrams contributing to the tadpole residue to one-loop order.}
\label{figure1}
\end{figure}
\noindent
Therefore, through working at the one-loop level within the TRSM, in a complex dimensional regularisation $d_L=4-2/L=2$, the relevant contributions can be illustrated by the diagrams in Figure \ref{figure1}, where the straight, wiggly, short-dashed and long-dashed lines stand respectively for fermions, vector bosons, scalars, and Faddeev-Popov ghosts contributions. Consequently, the corresponding one loop residue is thus given by 
\begin{equation}
\Delta_1 = \sum_{i}\,V_i\,A(m_i^2)
\label{eq:residue}
\end{equation}
where the sum runs all over the previous contributions, $V_i$ includes all the vertices, Dirac traces as well as the symmetry factors and the $A(m_i^2)$, that unfolds as a simple form in terms of the Passarino-Veltman function \cite{Passarino:1978jh}, is a pure U.V. divergent term that can be factorized to provide reliable results. 

Although we do not provide a wide calculation we, for more details, refer the reader to more advanced discussions in the previous works \cite{Rahili:2015nel,Rahili:2018ert}. Also, we would like to clarify a very important point, that such cancellation does not depend upon symmetry breaking, and our results could be obtained so, in the symmetry unbroken phase.

\subsection{Phenomenological Analysis}
\label{subsec:pheno}
First, we mention that further investigation of the space parameter is done by interfacing the model with HiggsBounds-5.2.0beta \cite{Bechtle:2015pma} and HiggsSignals-2.2.0beta \cite{Bechtle:2013xfa} packages, thus testing the existing exclusion limits at 95$\%$ C.L. from Higgs searches at LEP, LHC, and Tevatron; checking the Higgs signal rate constraints from various searches and taking into account the recent LHC 13 TeV results in the framework of TRSM.

Accordingly, keeping only the dominant top quark contribution and neglecting leptons, the modified VC in the framework of TRSM for $h_1$ that should be canceled, or at least be kept at a manageable level, reads as: 
\begin{equation}
\begin{split}
\delta m_{h_1}^2\,\sim~& 2m_{W}^2+m_{Z}^2-3\,\it{Tr}(I_n)\,m_t^2+ \Big(3\lambda+\lambda_4+\lambda_5\Big)\vew^2
\end{split}
\label{eq:vc-h1}
\end{equation}
where $\it{Tr}(I_n)=2^{n/2}=2$ is the Dirac trace associated with the fermion loops.\\
If we consider no mixing between the SM doublet $H$ and the Singlet fields i.e. $S_1$ and $S_2$, which is achieved by simultaneously taking $\lambda_4,\lambda_5 \to 0$, then our result agrees with that of Ref. \cite{Veltman:1980mj,RuizAltaba:1989gu} and reads $$\frac{3}{2}h-\frac{3}{2}\it{Tr}(I_n)\,t+c^2+\frac{1}{2}=0,$$
with $h \equiv m_{h_1}^2/m_Z^2$, $t \equiv m_{t}^2/m_Z^2$ and $c^2 \equiv \cos^2\theta_w$. Even so, for the restricting of the allowed TRSM parameter space, approximate expression of $\lambda_4+\lambda_5$ from (\ref{eq:vc-h1}) could be used,
\begin{equation}
\lambda_4+\lambda_5 = (6\,m_t^2-2\,m_W^2-m_Z^2-3\,m_{h_1}^2)/\vew^2 \sim 1.85,
\label{ }
\end{equation}
in order to cancel the quadratic divergences.

\noindent
Similarly to the Higgs boson mass, $m_{h_1}$, the remaining scalar singlet masses are also shifted via two quadratically divergent corrections,
\begin{equation}
\begin{split}
\delta m_{h_2}^2\,\sim~& \Big( 3\lambda_1+\lambda_3+2\,\lambda_4 \Big)\vso^2 \\
\delta m_{h_3}^2\,\sim~& \Big( 3\lambda_2+\lambda_3+2\,\lambda_5 \Big)\vst^2,
\end{split}
\label{eq:vc-h2h3}
\end{equation}
thus laying a two new constraints to TRSM space parameters. The latter expressions reduce to $ \Big( 3\lambda_1+2\lambda_4 \Big)\vso^2$ and $\Big( 3\lambda_2+2\lambda_5 \Big)\vst^2$ in the limit of no mixing between the two singlets, which well match the results in \cite{Kundu:1994bs,Grzadkowski:2009mj,Bazzocchi:2012pp,Antipin:2013exa,Karahan:2014ola}. 

\noindent
Hereafter we provide a more comprehensive analysis of the VC requirement done in the TRSM framework.

\begin{figure}[!hb]
\hspace*{0.0cm}
\begin{minipage}{8cm}
\includegraphics[width=8cm,keepaspectratio=true]{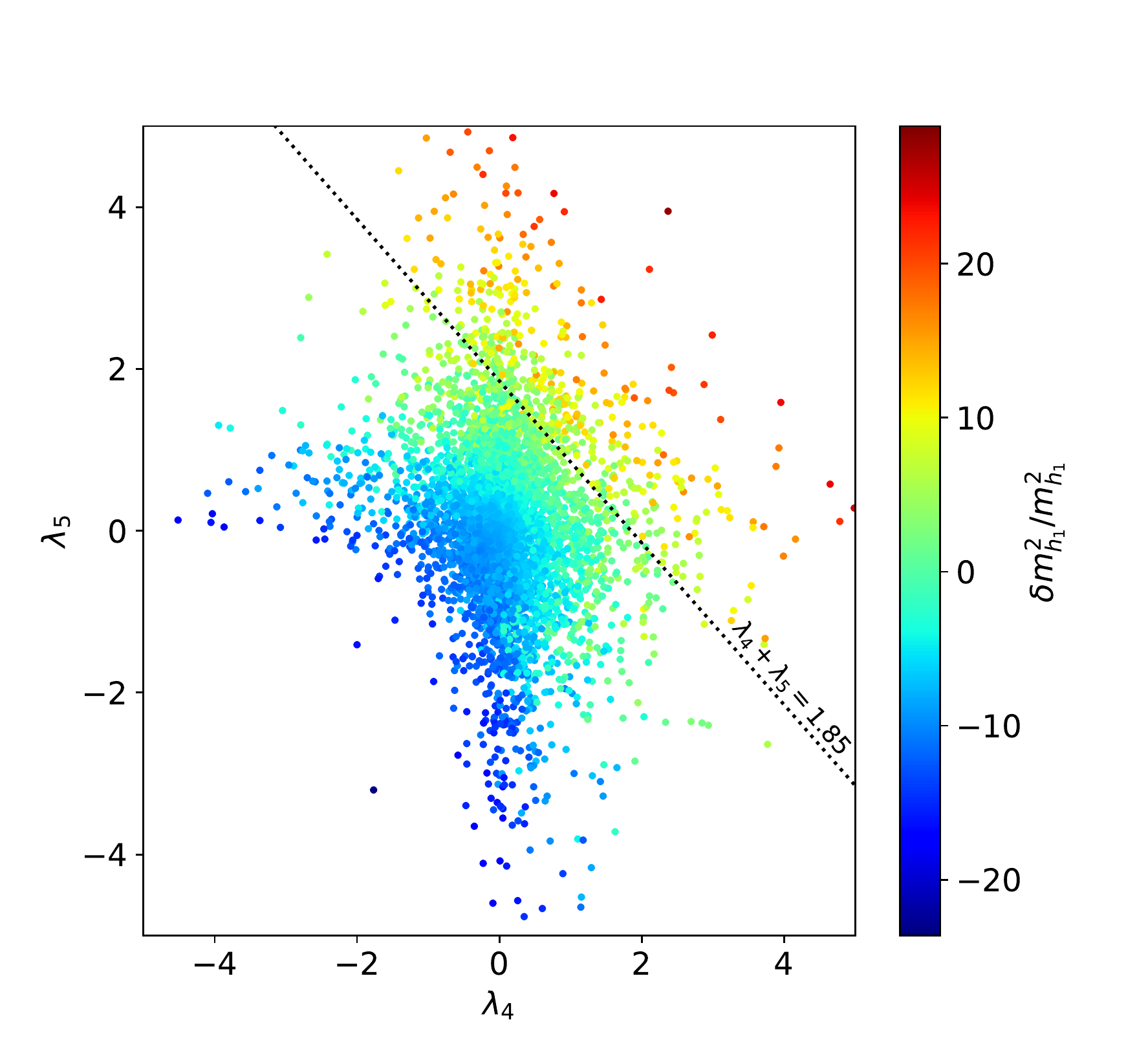}
\end{minipage}
\caption{Correction $\delta m_{h_1}^2/m_{h_1}^2$ in the TRSM shown as a scatter plot in the ($\lambda_4$,$\lambda_5$) plan.}
\label{figure2}
\end{figure}

Firstly and most crucially, the one-loop quadratically divergent correction to $m_{h_1}^2$, given by Eq.(\ref{eq:vc-h1}), sets a lower (upper) possible value for $\lambda_4,\lambda_5$ in order to cancel the divergencies. As illustrated in Figure \ref{figure2}, the lowest possible value for $\lambda_4$ is $\approx-1.4$ corresponds to an upper limit for $\lambda_5$ of the order of $\approx3.3$, and thereafter increases almost inversely with $\lambda_5$, ending at $\approx2.45$ for $\lambda_5\approx-0.68$. Furthermore, it is noteworthy that all remaining couplings, i.e. $\lambda$, $\lambda_1$ and $\lambda_2$ are not impacted due to VC constraint, except $\lambda_3$ whose allowed range is narrowed; we found it nowhere outside $[-0.57, 1.18]$ interval.

\begin{figure}[!h]
\hspace*{0.0cm}
\begin{minipage}{8cm}
\includegraphics[width=8cm,keepaspectratio=true]{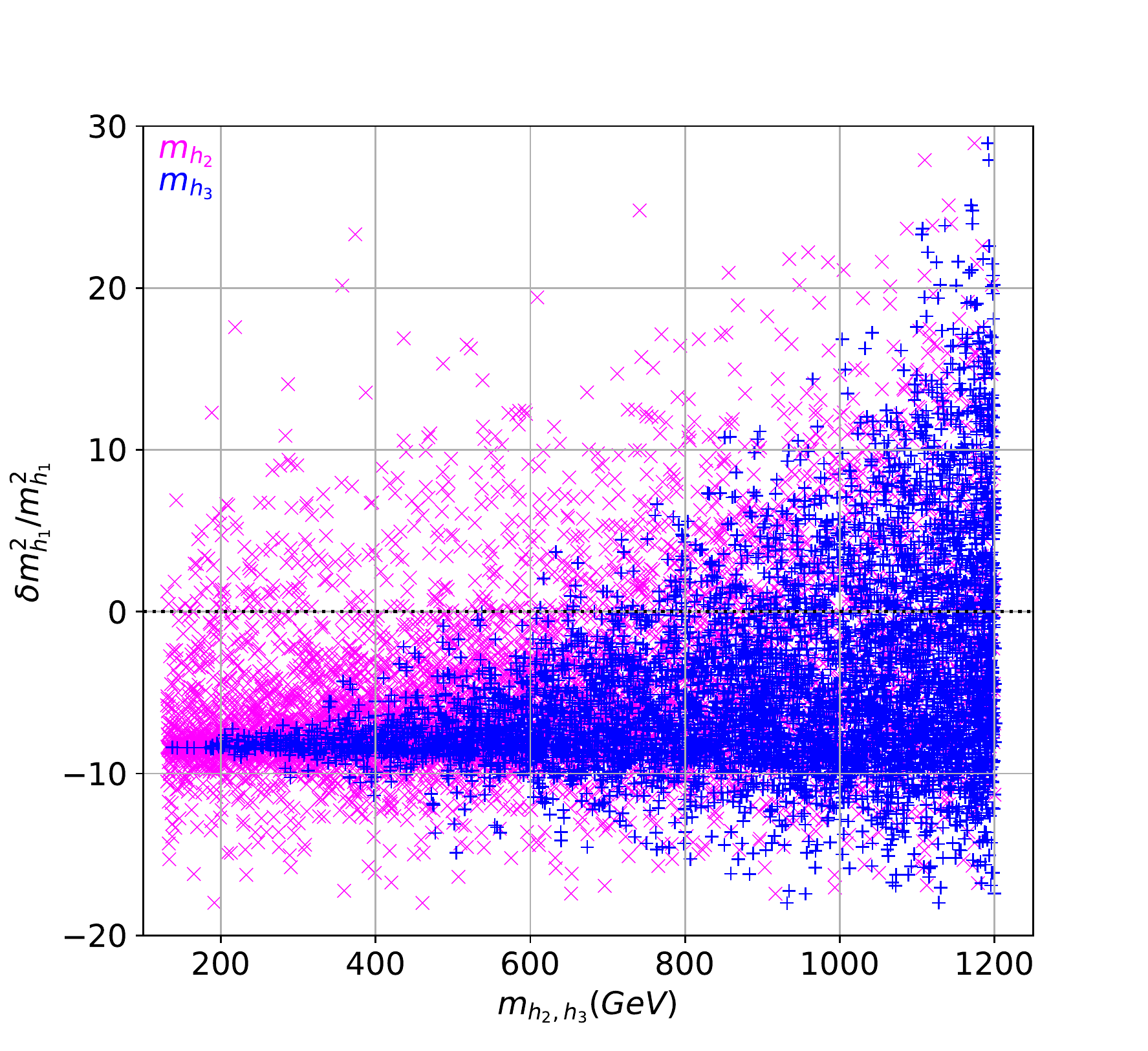}
\end{minipage}
\caption{Radiative correction $\delta m_{h_1}^2/m_{h_1}^2$ as a function of the Higgs bosons masses, $m_{h_2,h_3}$, in the TRSM.}
\label{figure3}
\end{figure}

Secondly, as for Higgs masses, one can notice that cancellation of quadratic divergencies sounds possible all over the $m_{h_2}$ range, which rendered it unaffected by the VC as can be seen in Figure \ref{figure3}, while for $m_{h_3}$ it may occur only over a minimal value. This is, in fact, a remarkable feature of effects of the VC on the scalar mass spectrum, notably the NP mass dependence on $\lambda_4+\lambda_5$ which is stronger for $m_{h_3}$ than for $m_{h_2}$; if the VC is to be fulfilled, the lower bound on $m_{h_3}$ must stretch out 500 GeV. Nevertheless, these overall resulting ranges are compatible with the LHC exclusion limits for heavy resonances decaying into a pair of Z/W gauge boson final states \cite{Aaboud:2018bun,Sirunyan:2018qlb}. 

\begin{figure}[!ht]
\hspace*{0.0cm}
\begin{minipage}{8cm}
\includegraphics[width=8cm,keepaspectratio=true]{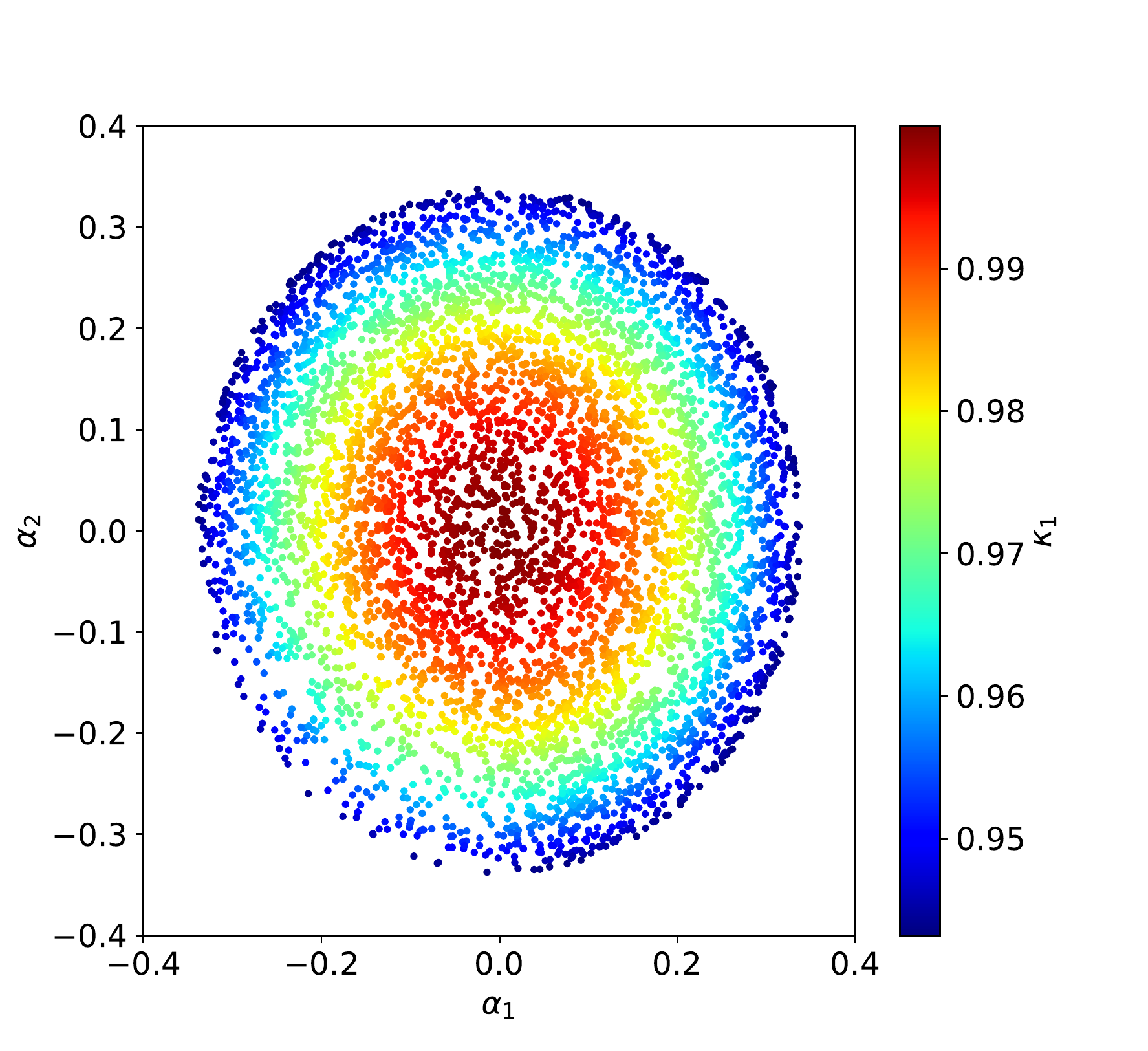}
\end{minipage}
\begin{minipage}{8cm}
\includegraphics[width=8cm,keepaspectratio=true]{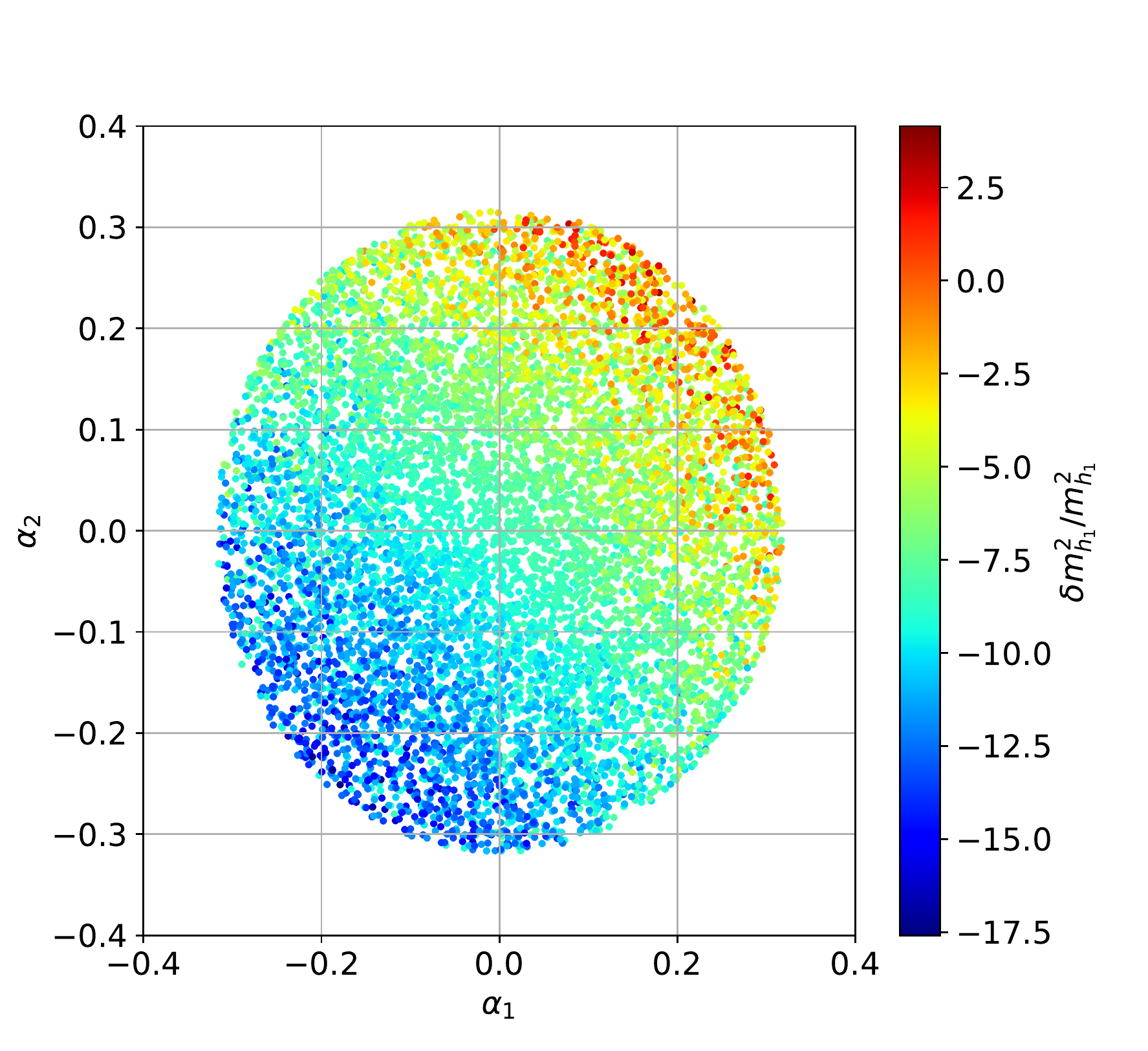}
\end{minipage}
\caption{$\mu_{ggF}$ signal strength (upper panel) and $\delta m_{h_1}^2/m_{h_1}^2$ radiative correction (lower panel) shown as a scatter plot in the ($\alpha_1$,$\alpha_2$) plan within the TRSM.}
\label{figure4}
\end{figure}

As with other parameters, we point out that realization of the Veltman condition can involve extensive scrutiny for the mixing angle $\alpha_i$ (i=1,2,3), which should afford theoretical insights for the best-fit signal strength of the $h_1$ for {\it ggF} production. The latter is directly related to $\alpha_1$ and $\alpha_2$ in the TRSM, and reads as,
\begin{equation}
\mu_{ggF}=\frac{\sigma_{ggF}}{\sigma_{ggF}^{SM}}=\kappa_1=\cos^2\alpha_1\cos^2\alpha_2
\label{eq:bestfit}
\end{equation}
while its updated experimental value is
\begin{equation}
\mu_{ggF}= 
\begin{cases}
1.04^{+0.09}_{-0.09},&\text{ATLAS\cite{ATLAS:2019slw}}\\ 
\\
1.22^{+0.14}_{-0.12}, & \text{CMS\cite{Sirunyan:2018koj}}
\end{cases}
\label{bestfitexp}
\end{equation}
In this regard, the upper panel in Figure \ref{figure4} shows that our theoretical predictions deviate significantly from the CMS experiment measurements, whereas for ATLAS, they are in line at $1\sigma$. On the other hand, regarding the gluonic decay, $h_1 \to gg$, all the amplitudes $A_{\frac{1}{2}}$ at lowest order for fermionic contributions are shifted by the same 
reduced couplings, $\cos\alpha_1\cos\alpha_2$. This is what justifies the fact that $\kappa_1$ would remain under unity all over the TRSM space parameter. As well, we notice here that 
the allowed range for mixings angles would be $-\pi/9 \le \alpha_1,\alpha_2 \le +\pi/9$ at $1\sigma$.

However, the naturalness criteria could drastically influence the situation. To clarify this point further, we exhibit in the lower panel of Figure \ref{figure4} the prevalence of $\delta m_{h_1}^2/m_{h_1}^2$ in the $[\alpha_1,\alpha_2]$ plane. At first sight, contrary to the best-fit $\kappa$ that regains its standard value for $\alpha_1=\alpha_2=0$, no exact cancellation of the quadratic divergencies can be achieved in such alignment limit (the pure doublet compound $h_1$ remains unstable to ultraviolet corrections), even more so when both mixing angles are simultaneously varied within a negative interval. On the other side, the quadratic divergencies become less noticeable once both $\alpha_{1}$ and $\alpha_{2}$ become positive, and their cancellation restrict $\alpha_{1}$ and $\alpha_{2}$ to lie within more reduced intervals $0 \le \alpha_1 \le \pi/12$ and $0 \le \alpha_2 \le \pi/13$.

%

%pper panels

%
\begin{figure}[!h]
\hspace*{0.0cm}
\begin{minipage}{8cm}
\includegraphics[width=8cm,keepaspectratio=true]{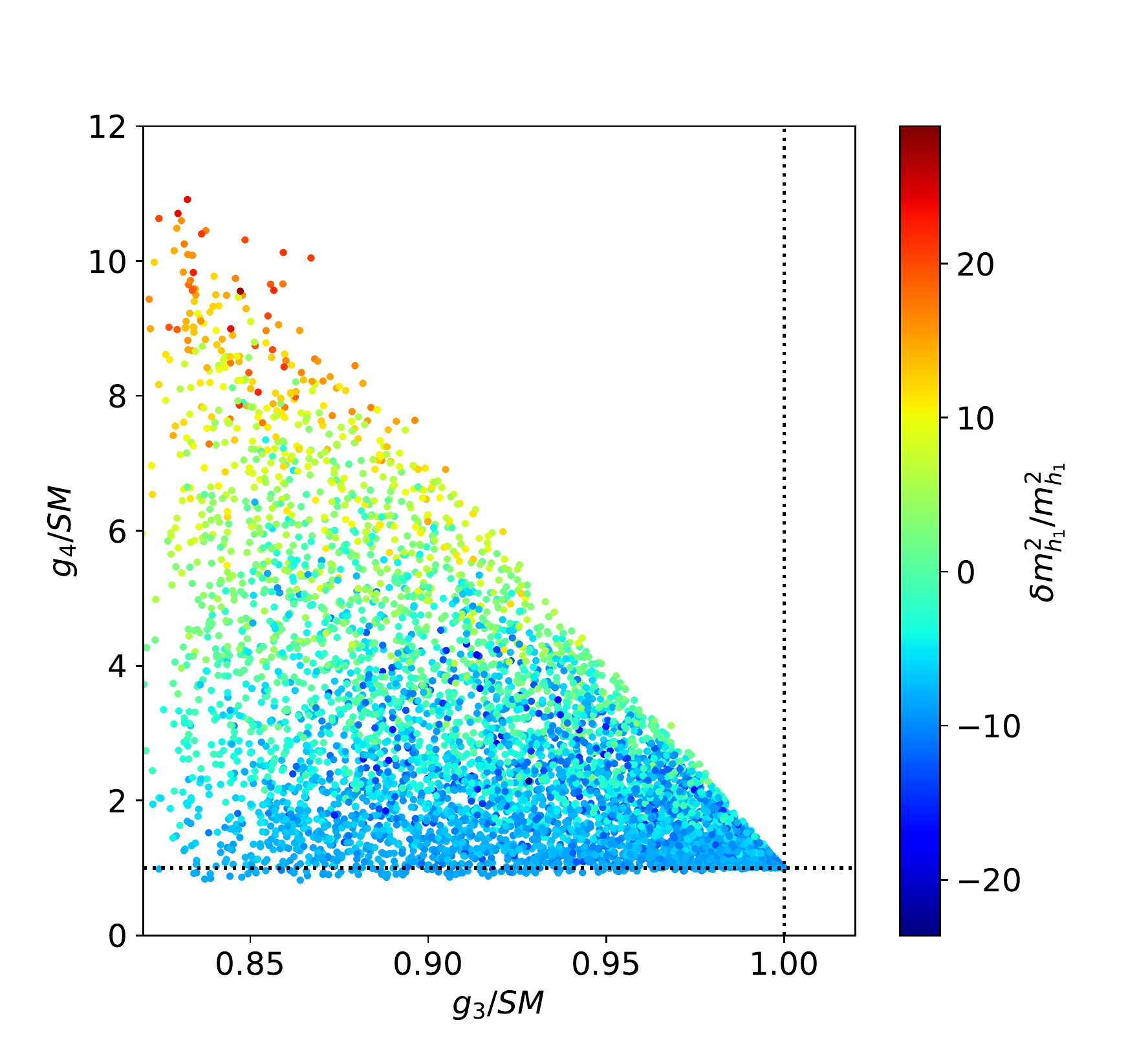}
\end{minipage}
\caption{$g_{3}/g_{3}^{SM}$ versus $g_{4}/g_{4}^{SM}$ in the TRSM. The palette color shows the radiative correction, $\delta m_{h_1}^2/m_{h_1}^2$.}
\label{figure5}
\end{figure}

Furthermore, the fact that the trilinear $g_3$ and quartic $g_4$ Higgs couplings are the most important that NP should nail down regarding their crucial role in ascertaining any extension BSM, the TRSM has provided a useful framework for addressing such purpose \cite{Jurciukonis:2018skr}. We, on our part, have investigated the impact of VC on these couplings that are expected to deviate from Standard Model. Figure \ref{figure5} below illustrates these couplings normalized to their SM value. A quick peek just before imposing VC shows that the $g_3$, which is positive, may not exceed more than $20\%$ below its SM value, whereas $g_4$ goes much further, reaching $\approx11\,g_4^{SM}$. 

However, once the VC conditions are turned on, the coupling $g_3$  decreased slightly ($0.84 \le g_{3}/g_{3}^{SM} \le 0.98$) yielding a significant reduction for the quartic coupling, e.g. $2 \le g_{4}/g_{4}^{SM} \le 6.1$. Nevertheless, this expected constraint on the latter is quite weak and it would therefore be appropriate to restrict it depending on specific values for $g_3$. For example, taking $g_3 \approx 0.96 g_{3}^{SM}$ required the $g_{4}/g_{4}^{SM} \in [2, 2.5]$, which, while it might not be a wide enough range, is compatible within $1\sigma$ level with recent indirect bounds on the quartic Higgs-self-couplings \cite{Borowka:2018pxx,DiMicco:2019ngk}. For the Higgs bosons masses, the VC seem to favor $m_{h_2}$ to roughly lie between the electroweak scale and 1 TeV as mentioned above, while for the heavy Higgs, $h_3$, must have a mass sufficiently high, e.g. $0.85\,{\rm TeV}\le m_{h_3}\le 1.15\, {\rm TeV}$, to impart the absence of fine-tuning in the Higgs mass parameter.

\section{Concluding Remarks}
\label{sec:conclusion}
In summary, the cancellation of quadratic divergences has been addressed within the TRSM, an extension of the SM with two extra real singlet scalar fields, $S_{1,2}$, giving rise to a wide phenomenology in seeking new physics. We have shown that at one loop and depending on the mixing angles, these quadratic divergencies are softened and go to zero within the allowed space parameter for the observed 125 GeV, but on the whole, the issue remains in its infancy as long as there is no allowed parameter space where the quadratic divergences simultaneously cancel out for both SM-like Higgs boson and the new scalar mass squares, increasingly pressing in light of recent LHC data. The tadpole equations for the new physics, as already derived, are not yet sufficient to make a critical test for quadratic divergencies cancellation; and therefore incorporation of either singlet or doublet vector fermions, that couple to fields $S_{1,2}$, will negatively affect the $\delta\,m_{h_2}^2$ and $\delta\,m_{h_3}^2$ to the extent that it doesn't enhance $\delta\,m_{h_1}^2$. This important point would seem to be the most plausible propositions to naturalize the three CP Higgs bosons all together.

\vspace{0.25cm}
\section{Acknowledgments}
The authors are grateful to Professors Abdesslam Arhrib and Rachid Benbrik for useful discussion and remarks.
This work is supported by the Moroccan Ministry of Higher Education and Scientific Research MESRSFC and CNRST: Project PPR/2015/6.

\end{document}